\documentclass[a4paper,11pt]{article}
\usepackage{geometry}
\usepackage{times}
\usepackage{graphics}
\usepackage{graphicx}
\usepackage{subfig}
\usepackage{amsfonts}
\usepackage{amsmath}
\usepackage{algorithm}
\usepackage{algorithmic}
\usepackage{amsthm,amssymb}
\usepackage[dvipsnames]{xcolor}

\usepackage[utf8]{inputenc}
\usepackage{listings}
\usepackage{listings-golang} %
\usepackage{listings-julia} %

\newcommand{\setlang}[1]{\lstset{ 
    language=#1 %
}}

\newcommand{\first}[1]{{\color{red} #1}}
\newcommand{\secd}[1]{{\color{blue} #1}}

\newcommand{\url}[1]{#1}

\newcommand{\sspace}{\renewcommand{\baselinestretch}{1.0}\large\normalsize}

\sspace

\usepackage[%
	pagebackref=true,
	pdfauthor={Pascal Fua, Krzysztof Lis},
	pdfsubject={Comparing Python, Go, and C++ on the $N$-Queens Problem},
	pdfcreator={},
	pdfproducer={},
	pdfkeywords={python, numba, go, C++, julia, n-queens, performance}, %
]{hyperref}

\title{Comparing Python, Go, and C++ on the $N$-Queens Problem}
\author{Pascal Fua, Krzysztof Lis\\Computer Vision Laboratory, EPFL}
\begin{document}

\maketitle

\begin{abstract}

Python currently is the dominant language in the field of Machine Learning but is often criticized for being slow to perform certain tasks. In this report, we use the well-known $N$-queens puzzle~\cite{Bell09} as a benchmark to show that once compiled using the Numba compiler it becomes competitive with C++ and Go in terms of execution speed while still allowing for very fast prototyping. This is true of both sequential and parallel programs.  In most cases that arise in an academic environment, it therefore makes sense to develop in ordinary Python, identify computational bottlenecks, and use Numba to remove them.

\end{abstract}

\begin{figure}[htbp]
\begin{tabular}{cccc}
\includegraphics[width=4cm]{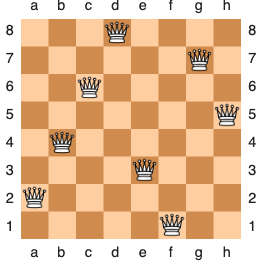}&\includegraphics[width=4cm]{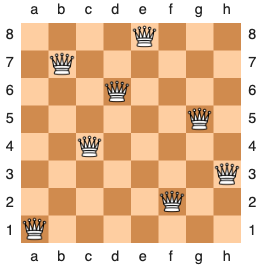}&\includegraphics[width=4cm]{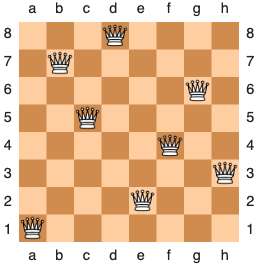}&\includegraphics[width=4cm]{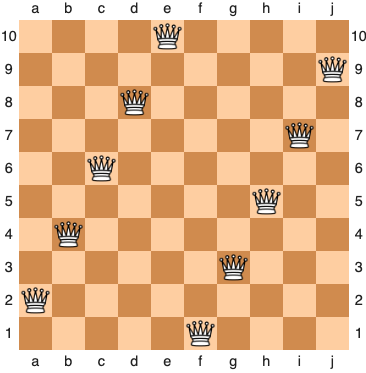}
\end{tabular}
\vspace{-6mm}
\caption{Three solutions of the puzzle on an $8 \times 8$ board and one on a $10 \times 10$ board.}
\label{fig:sols}
\end{figure}

\section{Introduction}

Python currently is the dominant language in the field of Machine Learning and gives easy access to powerful Deep Learning packages such as TensorFlow and PyTorch. However, it is known to be slow to perform some operations such as loops, which are not always easy to vectorize away. In such situations, one might consider switching to another language, such as C++ or the more recent Go language whose similarity to Python makes them potentially attractive replacements.  In this note, we will argue that this may  be necessary only rarely because the Numba python compiler~\cite{Numba} delivers performance close to those of C++ while preserving the compactness and ease of development that make Python such a powerful prototyping tool. Furthermore, it is easy to use. Once a set of Python functions has been identified as computationally intensive, one simply adds  Numba decorators before their definitions to instruct Python to compile them while leaving the rest of the code largely unchanged.

To demonstrate this, we use the well known $N$-queens puzzle~\cite{Bell09} as a benchmark. It involves placing $N$ chess queens on an $N \times N$ chessboard so that no two queens threaten each other. Fig.~\ref{fig:sols} depicts three solutions on a standard $8 \times 8$ board and one on a larger $10 \times 10$ one. We will focus on finding the number of solutions as a function of $N$. This is easy for small values of $N$ but quickly becomes computationally intractable for larger ones because the complexity of our algorithm is exponential with respect to $N$. There is no known formula for the exact number of solutions and, to date, $27$ is the larger value of $N$ for which the answer has been computed~\cite{Preusser09}. 

All our benchmarking code available is available online.\footnote{\url{https://github.com/cvlab-epfl/n-queens-benchmark}} We welcome comments and suggestions for potential improvements.

\section{Sequential Processing}

We started from the recursive algorithm described in~\cite{Wirth76} to compute one single solution of the 8-queens problem and translated it to Python. Our implementation relies on the fact that for two queens at board locations $(i_1,j_1)$ and $(i_2,j_2)$ not to be in conflict with each other, they must not be on the same row, on the same same column, or on the same diagonal. The first two mean that $i_1 \neq i_2$ and  $j_1 \neq j_2$. The third holds if $i_1+i_2  \neq j_1 + j_2$ and  $i_1-i_2  \neq j_1- j_2$. In other words, the two diagonals going through any $(i,j)$ location are completely characterized by $i+j$ and $i-j$. 

To exploit this, the function \lstinline{allQueensRec} of Tab.~\ref{fig:python} allocates boolean arrays  \lstinline{col}, \lstinline{dg1},  and \lstinline{dg2} to keep track on which columns and diagonals are still available to place a new queen on an $n \times n$ board. It then invokes the recursive function \lstinline{allQueensRecursive}. At recursion level $i$, for each $j$, it adds a queen at location $(i,j)$ if it is available, marks the column $j$ and the diagonals $i+j$ and $i-j$ as unavailable for additional queens, and calls itself for row $i+1$. The recursion ends in one of two ways. Either $i$ reaches $n$, meaning that all rows have been successfully filled, or no more queen can be added. If the first case, a counter is incremented. In the second case, nothing happens. In both cases, the program backtracks, undoes it earlier marking,  and continues until all solutions have been found. This process could be sped up by exploiting the symmetries of the $N$-queens problems. However, this is not required for benchmarking purposes an we chose not to do it to keep the code simple. We also chose to use the C++ and Go naming convention for functions and variables, that is, we use \lstinline{allQueensRec} instead of the more typical  \lstinline{all_queens_rec}, so that we can use the same names in all versions of the code we present. 

\begin{figure}[htbp]
\centering
\begin{tabular}{cc}
\includegraphics[width=0.45\linewidth]{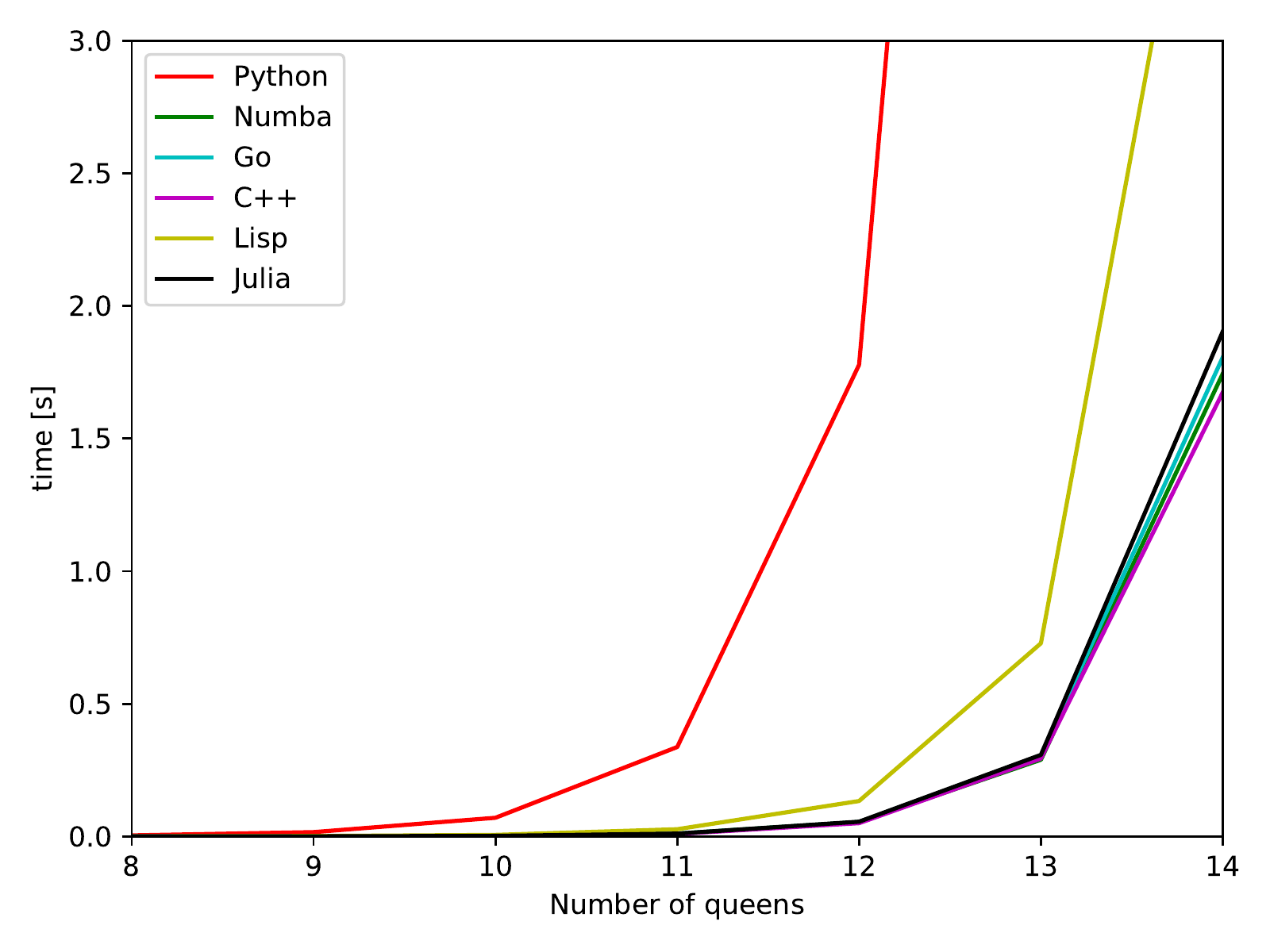}&\includegraphics[width=0.45\linewidth]{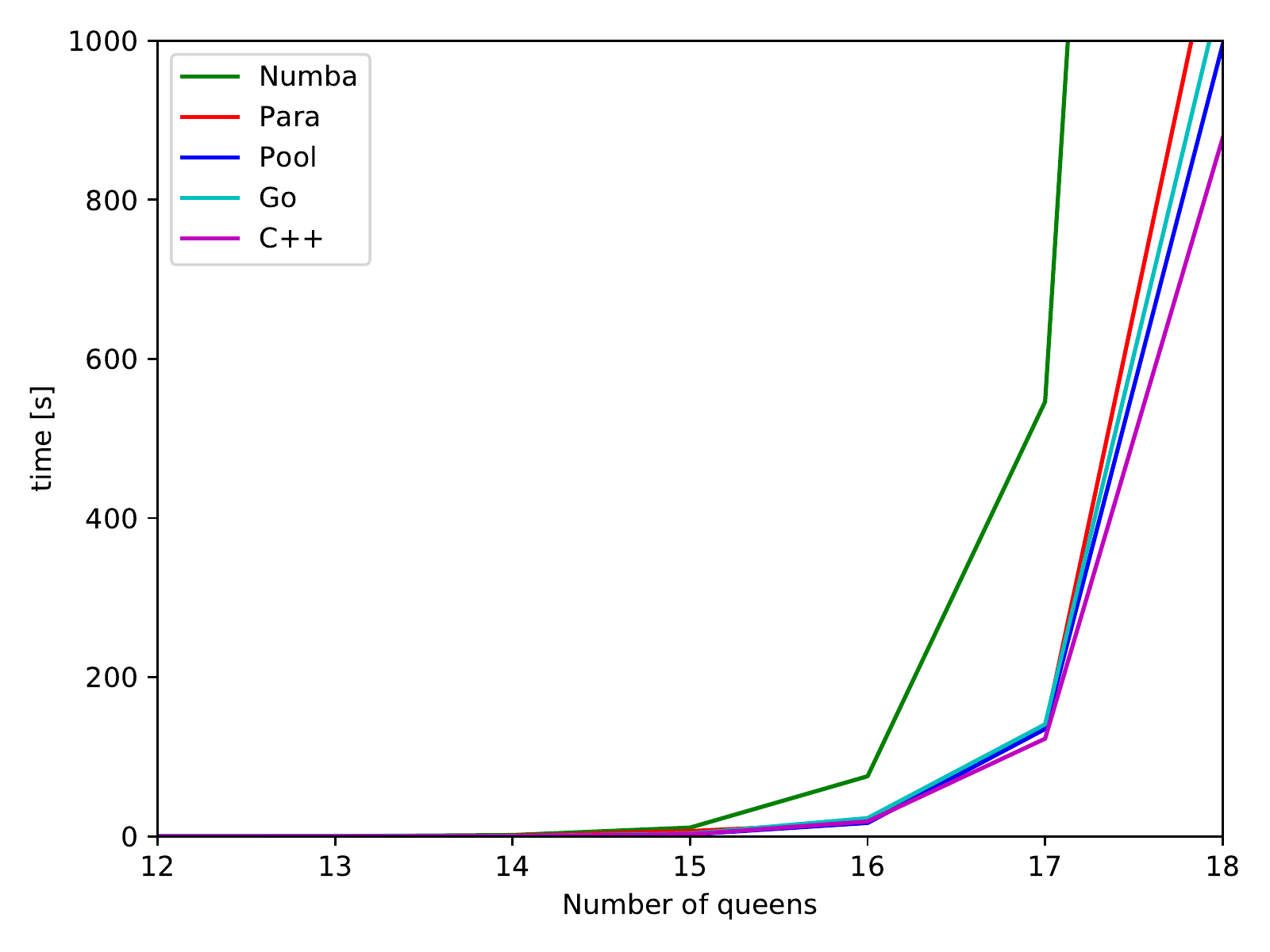}\\
\includegraphics[width=0.45\linewidth]{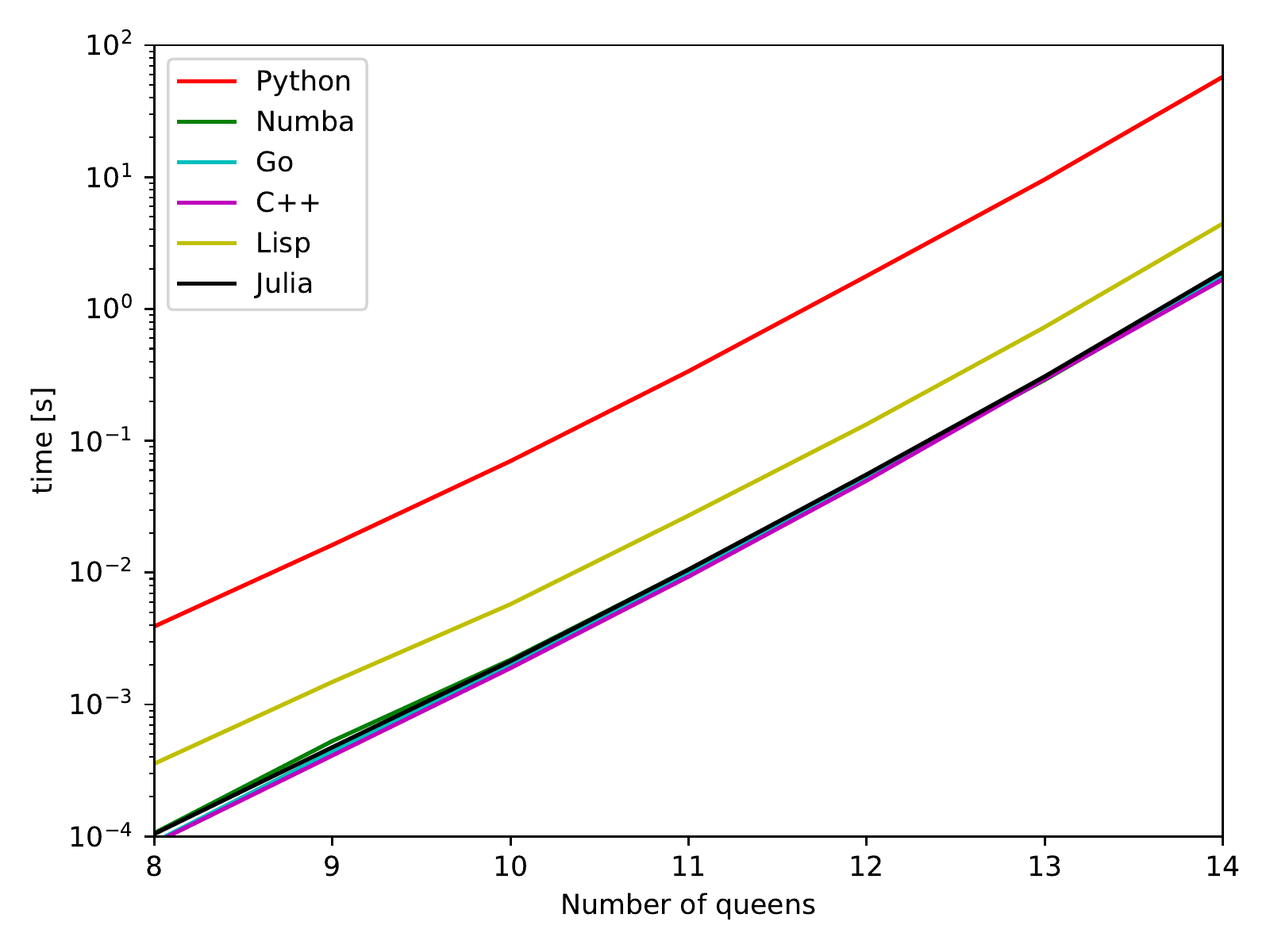}&\includegraphics[width=0.45\linewidth]{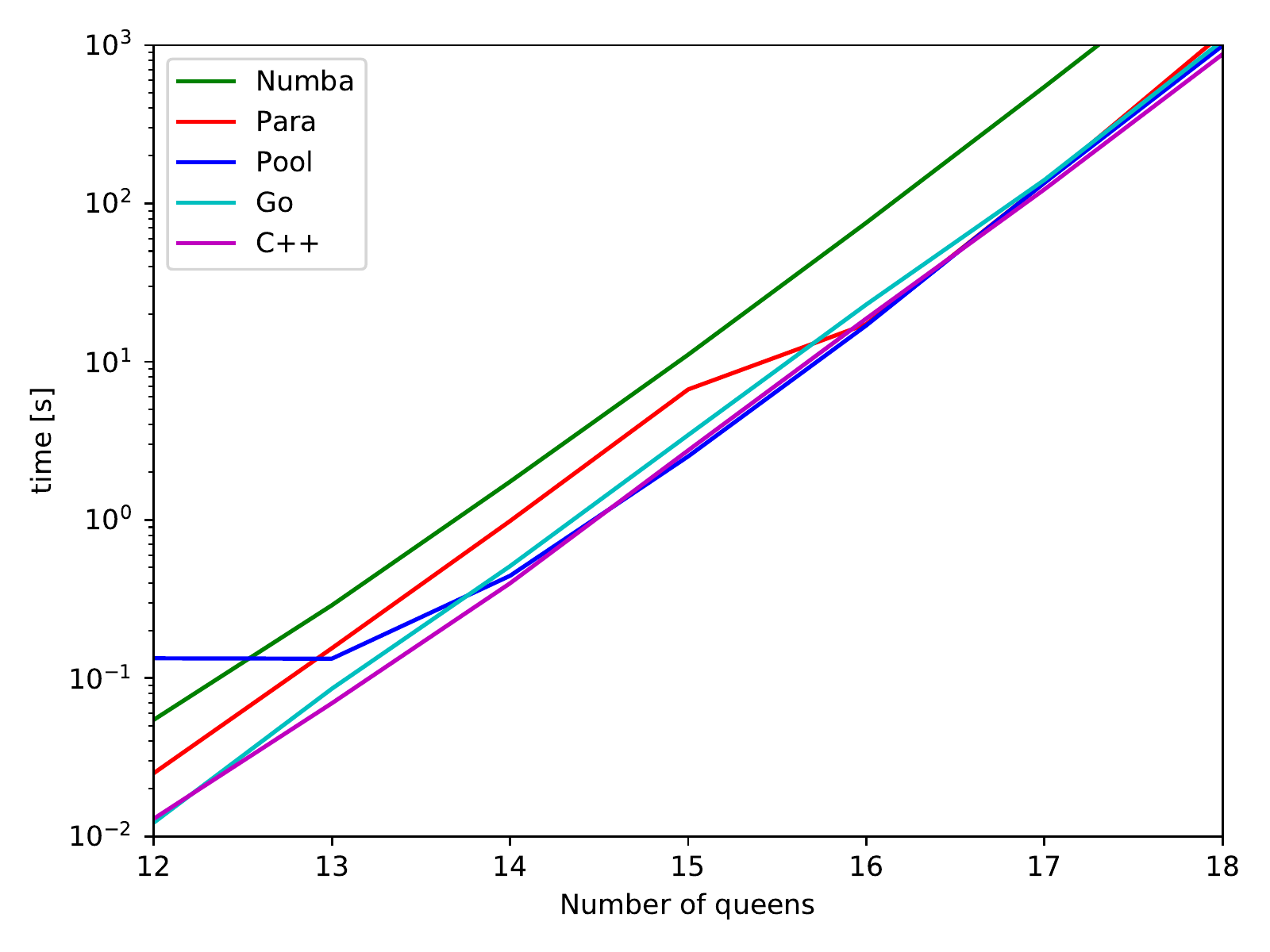}\\[-2mm]
(a)&(b)
\end{tabular}
\vspace{-4mm}
\caption{Run times as a function of the board size. Linear scale at the top and log scale at the bottom. (a) Sequential. (b) Parallel.}
\label{fig:seq}
\end{figure}

In Fig.~\ref{fig:seq}(a), the red curves depicts the computation time on a 2.9 GHz Quad-Core Intel Core i7 Mac running the Catalina operating system. In the top part of the figure, we plot the wall-clock time as as function of the board size $n$ using a standard scale. In the bottom part of the figure, we plot the same computation times using a log-scale instead, which results in an almost straight curve.  This serves as a sanity check because the computational complexity grows exponentially with $n$.  As  \lstinline{allQueensRecursive} performs loops, our vanilla Python implementation is inefficient. To remedy this, we used the Numba python compiler~\cite{Numba} as shown in Tab.~\ref{fig:numba}. The code is almost unchanged except for adding a couple of Numba decorators and  Yet, as depicted by the green curve in Fig.~\ref{fig:seq}(a), these minor modifications deliver a 33-fold increase in average computing speed of  \lstinline{allQueensNmb} over  \lstinline{allQueensRec}.

The Numba decorator \lstinline{njit()} that appears in Tab.~\ref{fig:numba} is short for \lstinline{jit(nopython=True)}. It ensures that if the code compile without errors it will {\it not} invoke python while running and will therefore be fast. Additionally, we could have used \lstinline{jit(nopython=True,nogil=True)} to instruct Numba to release the Python {\it Global Interpreter Lock}~\cite{GIL}  while executing the function, thus allowing several versions to run simultaneously on threads of a single process, something that standard Python code cannot do because of the aforementioned lock. This does not have any significant impact on performance in a sequential execution scenario.

\begin{table}
\begin{lstlisting}[language=Python]
def allQueensRec(n):   
    # Arrays used to flag available columns and diagonals
    col = np.ones(n,dtype=bool)
    dg1 = np.ones(2*n,dtype=bool)
    dg2 = np.ones(2*n,dtype=bool)
    
    return allQueensRecursive(n,0,col,dg1,dg2)
    
def allQueensRecursive(n,i,col,dg1,dg2):
    # All rows are filled, stop the recursion and report a new solution
    if n == i : 
        return 1
    # Try putting a queen in each cell of row i 
    nsol = 0
    for j in range(n):
        if (col[j] and dg1[i+j] and dg2[i-j+n]): 
            
            col[j]     = False  # Mark column j as occupied
            dg1[i+j]   = False  # Mark diagonal i+j as occupied
            dg2[i-j+n] = False  # Mark diagonal i-j as occupied

            nsol+=allQueensRecursive(n,i+1,col,dg1,dg2)
            
            col[j]     = True   # Unmark column j
            dg1[i+j]   = True   # Unmark diagonal i+j
            dg2[i-j+n] = True   # Unmark diagonal i-j
            
    return nsol
\end{lstlisting}
\vspace{-4mm}
\caption{Vanilla python code.}
\label{fig:python}
\end{table}

\setlang{Python}

\begin{table}
\begin{lstlisting}
@njit() #Numba decorator
def allQueensNmb(n,i=0,col=None,dg1=None,dg2=None):
    # np.bool_ not np.bool because of https://github.com/numba/numba/issues/1311   
    col = np.ones(n,dtype=np.bool_)
    dg1 = np.ones(2*n,dtype=np.bool_)
    dg2 = np.ones(2*n,dtype=np.bool_)

    return allQueensNumba(n,0,col,dg1,dg2)
    
@njit() #Numba decorator
def allQueensNumba(n,i,col,dg1,dg2):
    # All rows are filled, stop the recursion and report a new solution
    if n == i :
        return 1
    # Try putting a queen in each cell of row i
    nsol=0
    for j in range(n):
        if (col[j] and dg1[i+j] and dg2[i-j+n]): 

            col[j]     = False  # Mark column j as occupied
            dg1[i+j]   = False  # Mark diagonal i+j as occupied
            dg2[i-j+n] = False  # Mark diagonal i-j as occupied

            nsol+=allQueensNumba(n,i+1,col,dg1,dg2)
            
            col[j]     = True   # Unmark column j
            dg1[i+j]   = True   # Unmark diagonal i+j
            dg2[i-j+n] = True   # Unmark diagonal i-j
            
    return nsol
\end{lstlisting}
\vspace{-4mm}
\caption{The python code of Tab.~\ref{fig:python} slightly modified to force numba compilation. .}
\label{fig:numba}
\end{table}

To further assess how effective the Python/Numba combination is, we rewrote the code in Go and C++, as shown in Tabs.~\ref{fig:go1} and~\ref{fig:cplus1}. The short variable declarations make the Go code very similar to the Python code while being statically typed. The C++ code is slightly more verbose and one must remember to deallocate what has been allocated because there is no garbage collector.  As shown in Fig.~\ref{fig:cvector}, this can be remedied by using more sophisticated containers such as the standard \lstinline{vectors}  of C++ that are automatically deallocated at the end of the scope of their definition. Note that we used  \lstinline{vector<uint8_t>} as the type for our boolean arrays instead of the apparently more natural  \lstinline{vector<bool>}. We did this because the latter packs the bits densely and has to perform binary arithmetic to extract the requested bit for each access because memory can only be addressed down to whole bytes. In other words, it reduces memory usage at the expense of increased computation. As we are interested in speed, it is therefore more effective to explicitly use bytes (\lstinline{uint8_t}) for our purpose. Nevertheless, we have verified that even when using byte vectors, the implementation of  Fig.~\ref{fig:cvector} incurs a small, but noticeable, performance decrease with respect to that of Tab.~\ref{fig:cplus1} and we therefore chose to stick with it, even though it is less elegant. In short, unlike Go, C++ gives the programmer great freedom to carry out tasks in many different ways but it takes a lot expertise to exploit it effectively and to avoid the many lurking pitfalls. 

For example, unlike Python and Go,  C++ does not automatically check that one does not write beyond the bounds of arrays. As a result, the buggy code of Tab.~\ref{fig:cbug} runs but returns nonsensical values. We unintentionally made this mistake while translating the code from Python and, even though this is a short program, it took us a while to spot it. Of course, we could have used a tool such as valgrind, which would have detected the error, but this is far less convenient than being given a runtime warning. 
By default Numba does not perform bounds checks but they can be enabled using the decorator \lstinline{njit(boundscheck=True)}, which can be useful while debugging. 

The cyan and purple curves of Fig.~\ref{fig:seq}(a) depict the corresponding runtimes. The Numba, Go, and C++ curves are almost superposed. Closer examination of the raw numbers give in Tab.~\ref{fig:data} in appendix show that C++ wins. Go in slower by about $6\%$ and Numba by $12\%$. Numba is slower mostly for low values of $n$, which suggests that the algorithm itself runs just as fast but that calling the Numba function from Python involves an overhead. While the observed differences are statistically significant based on the variances of the different runs, in our daily research practice, they are rarely large enough to justify giving up the development speed that Python provides and to contend with potential bugs such as the one discussed above. 

However, there are optimizations that require the low-level control that C++ or Go can provide. For example, in all versions of the code presented here, the memory for the \lstinline{col}, \lstinline{dg1}, and \lstinline{dg2} arrays is allocated dynamically on the heap. The array sizes are decided at runtime and this code could in theory handle the $N$-queens problem for any value of $n$. However the computational cost is exponential and any value of $n>32$ is wildly impractical. If we accept to limit ourselves to $n<=32$, we can use fixed-sized arrays allocated on the stack by declaring them as  \lstinline{var col[32]bool} in Go or \lstinline{ std::array<bool, 32>} in C++. Unlike in the case discussed above, using \lstinline{bool} instead \lstinline{uint8_t} has no adverse effect. We have checked that the C++ code modified in this manner delivers a  $20\%$ gain over Numba, instead of the earlier  $12\%$. Potential explanations are that  putting the arrays on the stack works better for the CPU cache or that the optimizer has an easier time reasoning about fixed-size stack arrays. In any event, this shows that C++, and Go, being closer to the hardware may be useful to fine-tune code under some circumstances.

Go can therefore be considered a promising alternative to both Python and C++ because its run-time checks make bugs such as the one of Fig.~\ref{fig:cbug} easy to detect and correct. Furthermore, it is almost as concise as a Python and a little faster than Numba. However, some of its design features make it unwieldy in the prototyping role. For example, insisting that {\it all} variables and packages declared in a file be used makes sense for production code but is unhelpful when groping for a solution to a research problem: Commenting out a particular line of code, can mean many modifications in the file, which are unnecessary until a final solution has been found. Similarly not providing a full-fledged class-system can be understood as a way to discourage the writing of hard-to-maintain spaghetti code, which is commendable in production mode but unnecessarily rigid in prototyping mode.

\begin{table}
\begin{lstlisting}[language=Golang]
func allQueensRec(n int) int {
	// Allocate arrays
	col := make([]bool, n, n)
	dg1 := make([]bool, 2*n, 2*n)
	dg2 := make([]bool, 2*n, 2*n)
	// All columns and diagonals are initially available
	for i := 0; i < n; i++ {
		col[i] = true
	}
	for i := 0; i < 2*n; i++ {
		dg1[i] = true
		dg2[i] = true
	}
	// Perform the recursive computation and return the results
	return allQueensRecursive(n, 0, col, dg1, dg2)
}

func allQueensRecursive(n int, i int, col [32]bool, dg1 [64]bool, dg2 [64]bool) int {
	if n == i {
		return 1
	}
	nsol := 0
	for j := 0; j < n; j++ {
		if col[j] && dg1[i+j] && dg2[i-j+n] {
			col[j]     = false
			dg1[i+j]   = false
			dg2[i-j+n] = false
			
			nsol += allQueensRecursive(n, i+1, col, dg1, dg2)
			
			col[j]     = true
			dg1[i+j]   = true
			dg2[i-j+n] = true
		}
	}
	return nsol
}
\end{lstlisting}
\vspace{-4mm}
\caption{Go version of the python code of Tab.~\ref{fig:python}.}
\label{fig:go1}
\end{table}

\begin{table}
\begin{lstlisting}[language=C++]
int allQueensRecursive(int n,int i,bool *col,bool *dg1,bool *dg2)
{
	if (n == i) {
		return 1;
	}
	int nsol = 0;
	for (int j = 0; j < n; j++) {
		if (col[j] && dg1[i+j] && dg2[i-j+n]) {
			col[j]     = false;
			dg1[i+j]   = false;
			dg2[i-j+n] = false;

			nsol += allQueensRecursive(n, i+1, col, dg1, dg2);

			col[j]     = true;
			dg1[i+j]   = true;
			dg2[i-j+n] = true;
		}
	}
	return nsol;
}

int allQueensRec(int n)
{
	// Allocate dynamic memory on the heap
	bool *col = new bool[n];
	bool *dg1 = new bool[2*n];
	bool *dg2 = new bool[2*n];
	// All columns and diagonals are initially available
	memset((void *)col,1,n*sizeof(bool));
	memset((void *)dg1,1,2*n*sizeof(bool));
	memset((void *)dg2,1,2*n*sizeof(bool));
	// Perform the recursive computation
	int nsol = allQueensRecursive(n,0, col, dg1, dg2);
	// No garbage collector, must deallocate to prevent memory leaks
	delete[] col;
	delete[] dg1;
	delete[] dg2;

	return nsol;
}
\end{lstlisting}
\vspace{-4mm}
\caption{C++ version of the python code of Tab.~\ref{fig:python}.  To initialize the arrays, we could have used loops as in the Go code of Tab.~\ref{fig:go1}. Instead we used the lower level instruction \lstinline{ memset}, which performs the same tasks without loops and can therefore be expected to be faster.}
\label{fig:cplus1}
\end{table}

\begin{table}
\begin{lstlisting}[language=C++]
typedef vector<uint8_t> BoolArray;  // Use uint8_t instead of bool to boost efficiency 

int allQueensRecursive(int n,int i,BoolArray& col,BoolArray& dg1,BoolArray& dg2){
       ........
}

int allQueensRec(int n)
{
	BoolArray col(n,   true);
	BoolArray dg1(2*n, true);
	BoolArray dg2(2*n, true);

	int nsol = allQueensRecursive(n,0,col,dg1,dg2);

	return nsol;
}
\end{lstlisting}
\vspace{-4mm}
\caption{Using C++ vectors makes it unnecessary to explicitly free them. The call to allQueensRecursive has been slightly modified slightly so that they are passed by value instead of by reference and therefore not copied.}
\label{fig:cvector}
\end{table}

\begin{table}
\begin{lstlisting}[language=C++]
int allQueensRec(int n)
{
	// dg1 and dg2 are of size n instead of 2n
	bool *col = new bool[n];
	bool *dg1 = new bool[n];
	bool *dg2 = new bool[n];
	memset((void *)col,1,n*sizeof(bool));
	memset((void *)dg1,1,n*sizeof(bool));
	memset((void *)dg2,1,n*sizeof(bool));

        ........
}
\end{lstlisting}
\vspace{-4mm}
\caption{Buggy version of the C++ code of Tab.~\ref{fig:cplus1}. It runs but returns nonsensical results.}
\label{fig:cbug}
\end{table}

\section{Parallel Processing}

\begin{table}
\begin{lstlisting}[language=Python]
@njit()   
def allQueensCol(n,j):

    col = np.ones(n,dtype=np.bool_)
    dg1 = np.ones(2*n,dtype=np.bool_)
    dg2 = np.ones(2*n,dtype=np.bool_)
    # Put a queen in cell j of the first row
    col[j]   = False
    dg1[j]   = False
    dg2[n-j] = False
    # Fills the rest of the board starting with the second row
    return allQueensNumba(n,1,col,dg1,dg2)
    
if __name__ == "__main__":
    nsol = 0
    for j in range(8):
        nsol += allQueensCol(8,j)
\end{lstlisting}
\vspace{-4mm}
\caption{The python code of Tab.~\ref{fig:numba} rewritten to perform $n$ independent computations.}
\label{fig:par1}
\end{table}

\begin{table}
\begin{lstlisting}[language=Python]
@njit(parallel=True)
def allQueensPara(n):
    nsol = 0
    for j in prange(n):         # prange is only applicable inside jit(parallel=True)
        nsol+=allQueensCol(n,j)
    return nsol

def allQueensPool(n,np=None):
     with Pool_proc() as pool:  # Create a pool of processes
         nsols= pool.map(partial(poolWorker,n),range(n))
         return (sum(nsols))
 
def poolWorker(n,j):
    return allQueensCol(n,j)
\end{lstlisting}
\vspace{-4mm}
\caption{Two different ways to Invoke the function {\it allQueensCol} of Tab.~\ref{fig:par1} so that the computation is split into $n$ tasks potentially running on different cores. Note how compact this code is.}
\label{fig:par2}
\end{table}

\begin{table}
\begin{lstlisting}[language=Golang]
func allQueensPara(nd int) int {
    // Create the structure that will be used to synchronize
	var wg sync.WaitGroup
	wg.Add(nd)
    // Explicitly allow go to run on 8 cores
	runtime.GOMAXPROCS(8)

	sols := make([]int, nd)

	f := func(wg *sync.WaitGroup, n int, j int) {
		sols[j] = allQueensCol(n, j) // Result for a queen in cell k of first row
		wg.Done()                    // Flag the thread as complete
	}
	for j := 0; j < nd; j++ {
		go f(&wg, nd, j)             // Launch a new thread for each computation
	}
	wg.Wait()                        // Wait for all threads to be completed 

	nsol := sols[0]                  // Sum the individual results
	for j := 1; j < nd; j++ {
		nsol += sols[j]
	}
	return nsol
}

func allQueensCol(n int, j int) int {

	col := make([]bool, n, n)
	dg1 := make([]bool, 2*n, 2*n)
	dg2 := make([]bool, 2*n, 2*n)

	for i := 0; i < n; i++ {
		col[i] = true
	}
	for i := 0; i < 2*n; i++ {
		dg1[i] = true
		dg2[i] = true
	}
	col[j]   = false
	dg1[j]   = false
	dg2[n-j] = false

	return allQueensRecursive(n, 1, col, dg1, dg2, 0)
}
\end{lstlisting}
\vspace{-4mm}
\caption{Go version of the parallel python code of Tabs.~\ref{fig:par1} and~\ref{fig:par2}.}
\label{fig:go2}
\end{table}

\begin{table}
\begin{lstlisting}[language=C++]
int allQueensCol(int n,int j) {

	bool *col = new bool[n];
	bool *dg1 = new bool[2*n];
	bool *dg2 = new bool[2*n];
	memset((void *)col,1,n*sizeof(bool));
	memset((void *)dg1,1,2*n*sizeof(bool));
	memset((void *)dg2,1,2*n*sizeof(bool));

	col[j]   = false;
	dg1[j]   = false;
	dg2[n-j] = false;

	int ncol = allQueensRecursive(n,1,col, dg1, dg2);

	free(col);
	free(dg1);
	free(dg2);
	return ncol;
}

int allQueensPara(int nd){

	vector<future<int>> running_tasks;
    // Start one process per column
	for(int col = 0; col < nd; col++){
		running_tasks.push_back(
			async(std::launch::async, [=]() {return allQueensCol(nd,col);})
		);
	}
	// Wait for results
	int nsol_sum = 0;
	for(auto& f : running_tasks) {
		nsol_sum += f.get();
	}
	return nsol_sum;
}
\end{lstlisting}
\vspace{-4mm}
\caption{C++ version of the parallel python code of Tabs.~\ref{fig:par1} and~\ref{fig:par2}. The {\it async} template function makes the code very compact.}
\label{fig:cplus2}
\end{table}

Nearly every modern computer, including the one we used, has a multicore CPU and we can speed things up by running independent parts of the computation simultaneously on separate cores. In Go and C++, this can be done using multiple threads. Standard Python cannot do this due to the {\it Global Interpreter Lock}  (GIL)~\cite{GIL} that we have already encountered in the previous section. Fortunately, there are several workarounds and we explored two of them:
\begin{enumerate}

\item Using Numba's automatic parallelization~\cite{Numba_parallel}. Numba implements the ability to run loops in parallel as in Open Multi-Processing (OpenMP). The loop's body is scheduled in separate threads and the system automatically takes care of data privatization and reduction.

\item Using a pool of processes.  The pool distributes the computation into separate processes and tasks are sent to the available processors using a FIFO scheduling. Each process has its own interpreter and GIL, so they do not interfere. The price to pay is that objects need to be serialized and sent to the processes.

\end{enumerate}
To test these two approaches, we parallelized the \lstinline{allQueensRec} in a simple way. As shown in Tab.~\ref{fig:par1},  we defined a new function {\it allQueensCol} that puts a queen in column $j$ of the first row and then invokes the function  \lstinline{allQueensNumba} defined in Tab.~\ref{fig:numba} starting at the second row instead of the first, as in \lstinline{allQueensRec}. Summing the results for all possible values of $j$ yields the same results by performing $n$ independent computations. In Tab.~\ref{fig:par2}, we integrate this code into two functions that spread the tasks across separate cores: \lstinline{allQueensPara} uses the first method described above while \lstinline{allQueensPool} uses the second. We will refer to them as {\it para} and {\it pool} respectively.    Numba can take parallelization even further and produce functions that exploit the GPU. However, we did not explore this aspect in this study because our problem is not conducive to GPU processing. 

In Fig.~\ref{fig:seq}(b), we compare runtimes of the sequential Numba-compiled code  of the previous section with our two parallelized versions. As before, the sequential code is depicted by the green curve while the two parallel versions are depicted by the red and blue curves, labeled para and pool respectively.  {\it para} clearly delivers a significant improvement. However, for smaller values of $n$, we noted that {\it para} does not always fully use the 8 cores of our machines, which impacts its performance. For values of $n$ up to 13, the overhead  involved in spawning new processes dominates the computational cost of {\it pool} and makes it uncompetitive. However, for $n>13$, this overhead becomes negligible with respect to the rest of the computation and  {\it pool} starts dominating, albeit only by a small margin for large values of $n$, as can be seen in Tab.~\ref{fig:data}.  

To again compare against Go and C++, we rewrote the code in these two languages using their built-in multi-threading capabilities, as shown  in Tabs~\ref{fig:go2} and~\ref{fig:cplus2}. Note that we used the template function {\it std::async} function to make the C++ version compact.   The corresponding performance measurements are depicted by the cyan and purple curves of Fig.~\ref{fig:seq}(b). As before for small values of $n$, {\it para} and {\it pool} are uncompetitive because the initial overhead is too large. However, for larger values of $n$ they catch up and eventually do better than Go and almost as well as C++. In short, there are corner cases in which it might pay to switch from Python to C++ or go but it is not clear how pervasive they are  in our research practice.

\section{Numba Limitations}

In the two previous sections, we have argued that Numba is a powerful tool to painlessly compile potentially slow Python code so that it runs almost as fast as Go and C++. However, it also has limitations: Only a subset of  Python~\cite{Numba_supported_python} and NumPy~\cite{Numba_supported_numpy} features are available inside compiled functions. Numba has a  compilation mode that generates code able to handle all values as Python objects and uses the Python C API to perform all operations on such objects. Unfortunately, relying on this mode results in almost no performance gain over non-compiled code. This is why we used the \lstinline{njit()} decorator in all our examples. It yields much faster code but requires that the native types of all values in the function can be inferred, which is not necessarily true in standard Python. Otherwise, the compilation fails. 

In practice, this imposes an additional workload on the programmer who has to figure out what parts of the code are computationally expensive, encapsulate them in separate functions, and make sure that these functions can be compiled using the no-python mode that  \lstinline{njit()}  enforces. This is probably why there are ongoing efforts to optimize whole Python programs such as PyPy~\cite{PyPy}. Unfortunately, the results are not always compatible with libraries utilizing the C API, such as those routinely used in the field of scientific computing. As discussed in appendix, Julia is a potential alternative to Python/Numba that supports both high performance scientific computing and fast prototyping, is compiled, and could eventually address this problem.

\section{Conclusion}

As Computer Vision and Machine Learning researchers, we primarily need a language that allows us to test and refine ideas quickly while giving us access to as many mathematical, image processing, and machine learning libraries as possible. The latter spares us the need to reinvent the wheel every time we want to try something new. Maintainability and ability to work in large teams are secondary considerations as our code often stops evolving once the PhD student or post-doctoral researcher who wrote it leaves our lab. Before that happens, we typically make it publicly available to demonstrate that the ideas we published in conference and journals truly work and, in the end, that is often its main function. 

Python fits that bill perfectly at the cost of being slow when performing operations such as loops. Fortunately, as we showed in this report, this shortcoming can be largely overcome by using the Numba compiler that delivers performance comparable to that of C++, which itself tends to be faster than Go. This suggests that a perfectly valid workflow is to first write and debug a program in ordinary Python; identify the computational bottlenecks; and use Numba to eliminate them. This will work most of the time. In the rare cases where it does not, we can rewrite the relevant section of the code in C++ and call it from Python, which can be achieved using Cython~\cite{Cython} or pybind11~\cite{pybind11}. Interestingly, this approach harkens to the standard way one used to work in the much older Common Lisp language, as discussed in the appendix.

\clearpage
\appendix
\section*{Appendix}

\section{Other Languages}

\begin{table}
\begin{lstlisting}[language=Pascal]
program eightqueen1(output);

var i : integer; q : boolean;
    a : array[ 1 .. 8] of boolean;
    b : array[ 2 .. 16] of boolean;
    c : array[ -7 .. 7] of boolean;
    x : array[ 1 .. 8] of integer;
 
procedure try( i : integer; var q : boolean);
    var j : integer;
    begin 
    j := 0;
    repeat 
        j := j + 1; 
        q := false;
        if a[ j] and b[ i + j] and c[ i - j] then
            begin 
            x[ i    ] := j;
            a[ j    ] := false; 
            b[ i + j] := false; 
            c[ i - j] := false;
            if i < 8 then
                begin
                try( i + 1, q);
                if not q then
                    begin 
                    a[ j]     := true; 
                    b[ i + j] := true; 
                    c[ i - j] := true;
                    end
                end 
            else 
                q := true
            end
    until q or (j = 8);
    end;
 
begin
for i :=  1 to  8 do a[ i] := true;
for i :=  2 to 16 do b[ i] := true;
for i := -7 to  7 do c[ i] := true;
try( 1, q);
if q then
    for i := 1 to 8 do write( x[ i]:4);
writeln
end.
\end{lstlisting}
\vspace{-4mm}
\caption{Pascal program by Niklaus Wirth in 1976. It finds one solution to the eight queens problem.}
\label{fig:pascal}
\end{table}

\begin{table}
\begin{lstlisting}[language=Prolog]
/* Use clpfd package to loop through all configurations until a feasible one is found */
n_queens(N, Qs) :-
        length(Qs, N),
        Qs ins 1..N,
        safe_queens(Qs).

/* Predicate is true if the configuration is feasible */
safe_queens([]).
safe_queens([Q|Qs]) :- safe_queens(Qs, Q, 1), safe_queens(Qs).
safe_queens([], _, _).
safe_queens([Q|Qs], Q0, D0) :-
        Q0 #\= Q,
        abs(Q0 - Q) #\= D0,
        D1 #= D0 + 1,
        safe_queens(Qs, Q0, D1).
 
 /* Example */     
 ?- n_queens(8, Qs), labeling([ff], Qs).
   Qs = [1, 5, 8, 6, 3, 7, 2, 4] ;
   Qs = [1, 6, 8, 3, 7, 4, 2, 5] .
\end{lstlisting}
\vspace{-4mm}
\caption{Prolog version of the Pascal code of Tab.~\ref{fig:pascal} from https://www.metalevel.at/queens/.}
\label{fig:prolog}
\end{table}

\begin{table}
\begin{lstlisting}[language=Lisp]
(defun allQueensRec(n)
  (declare (type fixnum n))

  (let ((col (make-array n       :initial-element t :element-type 'boolean))
        (dg1 (make-array (* 2 n) :initial-element t :element-type 'boolean))
        (dg2 (make-array (* 2 n) :initial-element t :element-type 'boolean)))
    (declare (type (array boolean 1) col dg1 dg2 ))
    
    (allQueensRecursive n 0 col dg1 dg2 0)))

(defun allQueensRecursive(n i col dg1 dg2)
  ;; Optional declarations. Some compilers exploit them to speed up the code
  (declare (type (array boolean 1) col dg1 dg2 ))
  (declare (type fixnum n i))

  (if (= i n)
      
      1

    (let ((nsol 0))
       (declare (type fixnum nsol))
    
      (loop for j from 0 below n 
            when (and (aref col j) (aref dg1 (+ i j)) (aref dg2 (- (+ i n) j)))
            do
              
              (setf 
                (aref col j) nil
                (aref dg1 (+ i j)) nil
                (aref dg2 (- (+ i n) j)) nil)
              
              (incf nsol (allQueensRecursive n (+ i 1) col dg1 dg2))
              
              (setf
                (aref col j) t
                (aref dg1 (+ i j)) t
                (aref dg2 (- (+ i n) j)) t))
      nsol)))
\end{lstlisting}
\vspace{-4mm}
\caption{Common Lisp version of the python code of Tab.~\ref{fig:python}.}
\label{fig:lisp}
\end{table}

\begin{table}
\begin{lstlisting}[language=julia]
function allQueensRecursive(n, i, col, dg1, dg2)
    # All rows are filled, stop the recursion and report a new solution
    if n == i
        return 1
    end

    # Try putting a queen in each cell of row i
    nsol = 0

    for j = 0:n-1
        if (col[j+1] && dg1[i+j+1] && dg2[i-j+n+1])

            col[j+1]     = false  # Mark column j as occupied
            dg1[i+j+1]   = false  # Mark diagonal i+j as occupied
            dg2[i-j+n+1] = false  # Mark diagonal i-j as occupied

            nsol += allQueensRecursive(n,i+1,col,dg1,dg2)
                    
            col[j+1]     = true   # Unmark column j
            dg1[i+j+1]   = true   # Unmark diagonal i+j
            dg2[i-j+n+1] = true   # Unmark diagonal i-j
        end
    end
    return nsol
end

function allQueensRec(n)
    col = ones(Bool, n)
    dg1 = ones(Bool, 2*n)
    dg2 = ones(Bool, 2*n)
    
    return allQueensRecursive(n,0,col,dg1,dg2)
end
\end{lstlisting}
\vspace{-4mm}
\caption{Julia version of the code of Tab.~\ref{fig:python}}
\label{fig:julia}
\end{table}

In his book~\cite{Wirth76}, N. Wirth proposed the $N$-queens algorithm implemented in Pascal and shown in  Tab.~\ref{fig:pascal}. It is specific to the case $n=8$ and is designed to stop when the first solution is found. It then returns the corresponding configuration. As shown in Tab.~\ref{fig:prolog}, this can be done even more concisely in Prolog, a logic programming language of the same era as Pascal. What makes Prolog particularly interesting is that, of all the languages discussed in this report, it is the only one that forces a truly different approach to programming. It is well suited to performing the kind of systematic exploration and backtracking that solving the  $N$-queens problem requires and is still used to solve specific tasks that involve rule-based logical queries such as searching databases.

As not all implementations of the Pascal standard support dynamic arrays, extending the program of Tab.~\ref{fig:pascal} that is specific to the $n=8$ case to the general case would require manually allocating memory using pointers, much as in C. This would not be necessary in the even older Common Lisp language, as shown in Tab.~\ref{fig:lisp}. Although among the most ancient languages still in regular use, Lisp offers many of the same amenities as Python, that is, dynamically allocated arrays, sophisticated loop structures, and garbage collection among others.  It can be used either in interpreted or compiled form, much like Python without and with Numba. When invoking the compiler, the optional type declarations help it generate faster code. A standard approach to developing in Lisp is therefore to prototype quickly without the declarations and then add them as needed to speed up the code. Interestingly Python is now moving in that direction with its support for type hints but does not yet enforce that variable and argument values match their declared types at runtime. They are only intended to be used by third party tools such as type checkers, IDEs, and linters~\cite{PythonTypes}.

As can be seen in Fig.~\ref{fig:seq}(a), because it is compiled, Lisp does not perform so badly compared to the more recent languages we discussed in this report. The even newer Julia~\cite{julia_lang} language can be understood as being related to it in that it supports rapid prototyping by allowing 
interactive execution while being compiled. Type declarations are available but optional and the code is very Python-like, as can be seen in 
Tab.~\ref{fig:julia}. Like Numba, Julia uses LLVM~\cite{LLVM} to perform low level optimizations and produce efficient native binary binary code. Unlike in C++ or Go, there is no explicit compilation step and yet it delivers performance that are almost on par with those of the other compiled languages we discussed, as can be seen in Fig.~\ref{fig:seq}(a) and Tab.~\ref{fig:data}. 

This make Julia a potentially attractive alternative to Python/Numba. Unfortunately, there are some significant obstacles to its adoption. First, it still is a new language.  Some important features remain experimental and the number of third-party libraries is limited, whereas Python gives access to a wealth of powerful libraries, such as the deep learning ones that have become absolutely central to our research activities.  Furthermore, it features some design choices that differentiate it from currently popular languages~\cite{julia_lang_differences}, such as 1-indexed arrays and multiple-dispatch instead of classes. Whether or not these choices are wise,  they make it harder to switch from an established language like Python to Julia.

\begin{table}
\begin{footnotesize}
\begin{tabular}{|r|c|c|c|c|c|c|c|c|c|c|c|}
\hline{\bf Sequential}& 8 & 9 & 10 & 11 & 12 & 13 & 14 & 15 & 16 & 17 & 18 \\
\hline
Python & 0.00391 & 0.01625 & 0.07024 & 0.33707 & 1.77808 & 9.58502& 57.5610 & - & - & - & - \\
\hline
Lisp & 0.00036 & 0.00148 & 0.00577 & 0.02711 & 0.13346 & 0.72808 & 4.43882 & 29.1243 & - & - & - \\
\hline
Julia & 0.00010 & 0.00047 & 0.00213 & 0.01057 & 0.05555 & 0.30745 & 1.90335 & 13.1295 & 89.8649 & 634.371 & - \\
\hline
Numba & 0.00011 & 0.00053 & 0.00219 & 0.01054 & 0.05455 & \secd{0.28898} & \secd{1.74438}& \secd{11.0766} &  75.5948 & 546.312 & 4085.54 \\
\hline
Go & \secd{0.00009} & \secd{0.00044} & \secd{0.00198} & \secd{0.01007} & \secd{0.05404} & 0.30519 & 1.80617 & 10.9284 & \secd{73.6867} & \secd{527.514} & \secd{4014.05} \\
\hline
C++ & \first{0.00009} & \first{0.00041} & \first{0.00189} & \first{0.00935} & \first{0.04991} & \first{0.29386} & \first{1.67399} & \first{9.99408} & \first{66.7773} & \first{491.910} & \first{3677.10} \\
\hline{\bf Parallel}& 8 & 9 & 10 & 11 & 12 & 13 & 14 & 15 & 16 & 17 & 18 \\
\hline
Numba Para & 0.00008 & 0.000248 & 0.00087 & 0.00473 & 0.02509 & 0.15463 & 0.98384 & 6.6811 & 17.2807 & 135.166 & 1184.33 \\
\hline
Numba Pool & 0.13120 & 0.130002 & 0.13113 & 0.13105 & 0.13349 & 0.13280 & \secd{0.44365} & \secd{2.52153} & \secd{16.9606} & \secd{135.078} & \secd{994.799} \\
\hline
Go &  \first{0.00012} &  \secd{0.000221} &  \first{0.00055 }&  \first{0.00236} &  \first{0.01222} & \secd{0.08580} & 0.51022 & 3.44080 & 22.9841 & 140.860 & 1068.71 \\
\hline
C++ & \secd{0.00012} & \first{0.000202} & \secd{0.00063} & \secd{0.00272} &  \secd{0.01293} & \first{0.06945} & \first{0.39773} & \first{2.75142} & \first{18.7342} & \first{122.827} &\first{ 877.398} \\
\hline
\end{tabular}
\end{footnotesize}
\vspace{-4mm}
\caption{Benchmarking results in seconds per trial for $n$ from 8 to 18.}
\label{fig:data}
\end{table}

\section{Raw Data}

The performance numbers we used to produce the plots of Fig.~\ref{fig:seq} are given in Tab.~\ref{fig:data}. For  both the sequential versions of the code and for each value of $n$, the time for the fastest implementation appears in red and the one for the second best in blue. These numbers were obtained on a 2.9 GHz Quad-Core Intel Core i7 Mac running the Catalina operating system. For all versions of the code we ran 20 trials for $8 \leq n \leq 15$, 10 for $16 \leq n \leq 17$, and 3 for $n=18$ and computed the mean and variance in each case. We have rerun all these benchmarks on an Intel Xeon X5690 CPU running Ubuntu 18.04 and the overall ranking of the implementations was unchanged.

\bibliographystyle{plain}
\bibliography{string,optim,misc}

\end{document}